\documentclass{aa}
\usepackage{amssymb}
\usepackage{latexsym}
\usepackage{graphicx}
\usepackage{longtable,lscape,graphicx}
\usepackage{amsmath}
\begin{document}


\title{{\lq Swing Absorption\rq} of fast magnetosonic waves\\
in inhomogeneous media}

\author{
B.M. Shergelashvili\inst{1}\fnmsep\thanks{On leave from the Center
for Plasma Astrophysics, Abastumani Astrophysical Observatory,
Kazbegi Ave. 2a, Tbilisi 380060, Georgia} \and T.V.
Zaqarashvili\inst{2} \and S. Poedts\inst{1} \and B.
Roberts\inst{3}}
\institute{Centre for Plasma Astrophysics, Katholieke
 Universiteit Leuven, Celestijnenlaan 200B, B-3001
Leuven, Belgium;
\email{Bidzina.Shergelashvili@wis.kuleuven.ac.be (BMS) \&
Stefaan.Poedts@wis.kuleuven.ac.be (SP)} \and Abastumani
Astrophysical Observatory, Kazbegi Ave. 2a, Tbilisi 380060,
Georgia; \\ \email{temuri@hubble.uib.es} \and School of
Mathematics and Statistics, University of St. Andrews, St.
Andrews, Fife, KY16 9SS, Scotland, U K;
\email{bernie@mcs.st-and.ac.uk}}
\offprints{B.M. Shergelashvili}

\date{Received 18 June 2004 / Accepted 23 August 2004}
\abstract{The recently suggested swing interaction between fast
magnetosonic and Alfv\'en waves (Zaqarashvili \& Roberts
\cite{paper1}) is generalized to inhomogeneous media. We show that
the fast magnetosonic waves propagating across an applied
non-uniform magnetic field can parametrically amplify the Alfv\'en
waves propagating along the field through the periodical variation
of the Alfv\'en speed. The resonant Alfv\'en waves have half the
frequency and the perpendicular velocity polarization of the fast
waves. The wavelengths of the resonant waves have different values
across the magnetic field, due to the inhomogeneity in the
Alfv\'en speed. Therefore, if the medium is bounded along the
magnetic field, then the harmonics of the Alfv\'en waves, which
satisfy the condition for onset of a standing pattern, have
stronger growth rates. In these regions the fast magnetosonic
waves can be strongly {\lq absorbed\rq}, their energy going in
transversal Alfv\'en waves. We refer to this phenomenon as {\lq
{\it Swing Absorption}\rq}. This mechanism can be of importance in
various astrophysical situations.
}
 \maketitle

\section{Introduction}
Wave motions play an important role in many astrophysical
phenomena. Magnetohydrodynamic (MHD) waves may transport momentum
and energy, resulting in heating and acceleration of an ambient
plasma. A variety of waves have recently been detected in the
solar atmosphere using the SOHO and TRACE spacecraft (see
Aschwanden 2003) and this has stimulated theoretical developments
(Roberts \cite{rob2}, \cite{rob1}; Roberts \& Nakariakov
\cite{robnak}). Hence, an understanding of the basic physical
mechanisms of excitation, damping and the interaction between the
different kinds of MHD wave modes is of increasing interest.
Formally speaking, there is a group of direct mechanisms of wave
excitation by external forces (e.g.\ turbulent convection,
explosive events in stellar atmospheres, etc.) and wave
dissipation due to non-adiabatic processes in a medium (such as
viscosity, thermal conduction, magnetic resistivity, etc.). There
is a separate group of wave amplification and damping processes
due to resonant mechanisms (see, for example, Goossens
\cite{goos}; Poedts \cite{poed}). This means that particular wave
modes may be damped (amplified) due to energy transfer
(extraction) into or from other kinds of oscillatory motions, even
when wave dissipation is excluded from consideration.

The interaction between different MHD wave modes may occur either
due to nonlinearity or inhomogeneity. The
nonlinear interaction between waves as resonant triplets
(multiplets) is well developed (e.g., Galeev \& Oraevski
\cite{galor}; Sagdeev \& Galeev \cite{saggal}; Oraevski
\cite{ora}). Moreover, the nonlinear interaction of two wave
harmonics in a medium with a steady background flow can also be
treated as a resonant triplet, with the steady flow being employed
as a particular wave mode with zero frequency (Craik
\cite{craik}). Other mechanisms of wave damping (or energy
transformation) are related to the spatial inhomogeneity of the
medium, including {\it phase mixing} (Heyvaerts \&
Priest \cite{heypr}; Nakariakov et al.\ \cite{nakrobmur}) and {\it
resonant absorption} (Ryutova \cite{ryu}; Ionson
\cite{ion}; Rae \& Roberts \cite{raerob}; Hollweg \cite{holl};
Poedts et al.\ \cite{poegok}; Ofman \& Davila \cite{ofm}). Due to
inhomogeneity, different regions of the medium have different
local frequencies. Resonant absorption arises in those regions where
the frequency of an incoming wave matches the local frequency of
the medium, i.e. where $\omega_i \approx \omega_l$, with $\omega_i$ the
frequency of the incoming wave and $\omega_l$ the local
frequency. The mechanical analogy of this process is the ordinary
mathematical pendulum forced by an external periodic action on the
pendulum mass. When the frequencies of the periodic external force
and the pendulum are the same the force can resonantly amplify the
pendulum oscillation.

An additional issue, linked to MHD wave coupling, is the
possibility of mutual wave transformations due to an inhomogeneous
background flow. In numerous studies it has been shown that the
non-modal (Kelvin \cite{kel}; Goldreich \& Linden-Bell
\cite{gollin}) temporal evolution of linear disturbances (due to
the wave number 'drift' phenomenon) brings different types of
perturbations into a state in which they satisfy the resonant
conditions so that wave transformations occur (e.g.\ see
Chagelishvili et al.\ \cite{charots}; Rogava et al.\
\cite{ropom}).

Recently, another kind of interaction between different MHD wave
modes, based on a parametric action, has been suggested
(Zaqarashvili \cite{zaq}; Zaqarashvili \& Roberts \cite{paper1},b;
Zaqarashvili, Oliver \& Ballester  \cite{zaq2,zaq3}). In this case
the mechanism of wave interaction originates from a basic physical
phenomenon known in classical mechanics as {\lq parametric
resonance\rq}, occurring when an external force (or oscillation)
amplifies the oscillation through a periodical variation of the
system's parameters. The mechanical analogy of this phenomenon is a
mathematical pendulum with periodically varying length. When the
period of the length variation is half the period of the pendulum
oscillation, the amplitude of the oscillation grows exponentially
in time. Such a mechanical system can consist of a pendulum
(transversal oscillations) with a spring (longitudinal
oscillations). A detailed description of such a mechanical system
is given in Zaqarashvili \& Roberts (\cite{paper1}) (hereinafter
referred to as Paper~I).

When an external oscillation causes the periodical variation of
either the wavelength (Zaqarashvili et al.\ \cite{zaq2,zaq3}) or
the phase velocity (Zaqarashvili \cite{zaq}; Zaqarashvili \&
Roberts \cite{zaq0} and Paper~I) of waves in the system, then the
waves with half the frequency of the external oscillations grow
exponentially in time. Energy transfer occurs between different
MHD wave modes in different situations. One specific case of wave
coupling takes place between fast magnetosonic waves propagating
across an applied magnetic field and Alfv\'en waves propagating
along this field. It has been shown that the standing fast
magnetosonic wave, which manifests itself through harmonic
variations of the density and the magnetic field, amplifies the
Alfv\'en waves with a perpendicular velocity polarization
(Paper~I). The frequency of the resonant Alfv\'en waves was found
to be half of the fast wave frequency. In Paper~I the ambient
medium was considered to be homogeneous. Therefore, the resonant
Alfv\'en waves had the same wavelength everywhere. However, if the
medium is inhomogeneous across the magnetic field, the wave length
of the resonant Alfv\'en waves will have different values in
different regions. If the medium is bounded along the magnetic
field lines, as, for example, in photospheric ~line-tying of
coronal magnetic field lines, then Alfv{\'e}n waves will be
amplified in specific regions only, viz.\ where the conditions for
the onset of a standing pattern are satisfied. In these regions
the fast waves can be strongly absorbed by Alfv{\'e}n waves with
half their frequency. To study this phenomenon, we first consider
fast magnetosonic waves propagating across a non-uniform magnetic
field. Then we study the absorption of these waves by
perpendicularly polarized Alfv{\'e}n waves propagating along the
field.
\section{Basic equations and equilibrium model}
\begin{figure*}
\begin{center}
\includegraphics[scale=0.8]{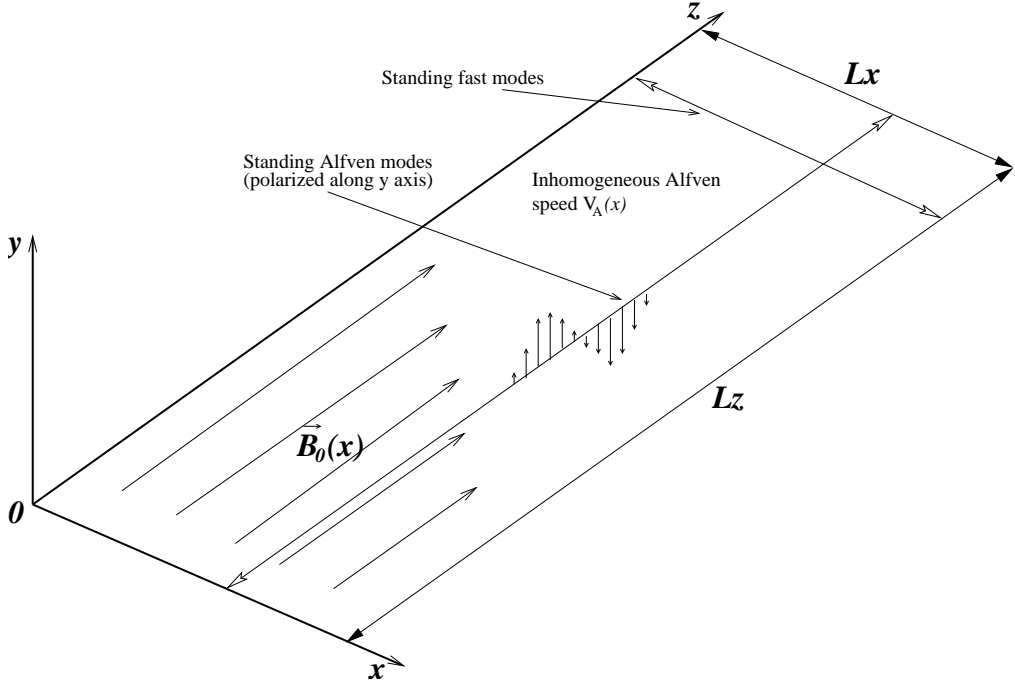}
\end{center}
\caption{Schematic view of the inhomogeneous background
configuration and the directions of wave propagation
(polarization).}\label{figbas}
\end{figure*}
Consider a magnetized medium with zero viscosity and infinite
conductivity, where processes are assumed to be adiabatic. Then
the macroscopic dynamical behaviour of this medium is governed by
the ideal magnetohydrodynamic (MHD) equations:
\begin{equation}\label{eqcont}
\frac{\partial \rho }{\partial t}+\vec{\nabla}\mathbf{\cdot }(\rho
\vec{U}%
)=0,
\end{equation}%
\begin{equation}\label{eqmot}
\rho \frac{\partial \vec{U}}{\partial t}+\rho \left( \vec{U}\cdot
\vec{\nabla%
}\right) \vec{U}=-\vec{\nabla}\left[ p+\frac{B^{2}}{8\pi }\right]
+\frac{%
\left( \vec{B}\cdot \vec{\nabla}\right) \vec{B}}{4\pi },
\end{equation}%
\begin{equation}\label{eqind}
\frac{\partial \vec{B}}{\partial t}+\left( \vec{U}\cdot
\vec{\nabla}\right)
\vec{B}=\left( \vec{B}\cdot \vec{\nabla}\right) \vec{U}-\vec{B}\left(
\vec{%
\nabla}\cdot \vec{U}\right) ,
\end{equation}%
\begin{equation}\label{eqener}
\frac{\partial p}{\partial t}+\left( \vec{U}\cdot \vec{\nabla}\right)
p=%
\frac{\gamma p}{\rho }\left( \frac{\partial \rho }{\partial t}+\left(
\vec{U}%
\cdot \vec{\nabla}\right) \rho \right) ,
\end{equation}
where $p$ and $\rho$ are the plasma pressure and density, $\vec U$
is the velocity, $\vec B$ is the magnetic field of strength $B=\vert
B \vert$ and $\gamma$
denotes the ratio of specific heats.

We consider an equilibrium magnetic field
directed along the $z$ axis of a Cartesian coordinate system,
$\vec{B}_{0}=(0,0,B_{0}(x))$. The equilibrium magnetic
field $\vec{B}_{0}$ and density $\rho _{0}$ are inhomogeneous in $x$.
The
force balance condition, from Eq.~(\ref{eqmot}), gives the
total (thermal + magnetic) pressure in the equilibrium to be
a constant:
\begin{equation}\label{balance}
p_{0}(x)+\frac{B^2_{0}(x)}{8\pi } = {\rm constant}.
\end{equation}
Also, the equation of state
\begin{equation}
p_{0}(x)= p_0(T_0(x),\rho_0(x)),
\end{equation}
relates the unperturbed pressure, density $\rho_0(x)$ and
temperature $T_0(x)$.

Eqs.~(1)-(4) are linearized around the static ($\vec{U_0}=0$)
equilibrium state (5). This
enables the study of the linear dynamics of magnetosonic and
Alfv\'en waves. A schematic view of the equilibrium configuration
is shown in Figure~\ref{figbas}.
\section{Fast magnetosonic waves}
Now let us consider the propagation of '{\lq pure\rq} fast magnetosonic
waves across the magnetic flux surfaces, i.e.\ along the $x$ axis,
taking {\bf $\vec u=(u_x,0,0)$} and $\partial/ \partial y \equiv 0$,
$\partial/
\partial z \equiv 0$. The linearization of equations
(\ref{eqcont})-(\ref{eqener}) then takes the form:
\begin{equation}\label{eqcontp}
\frac{\partial \rho _{1}}{\partial t}+u_{x}\frac{d \rho _{0}(x)}{d
x}+\rho _{0}(x)\frac{\partial u_{x}}{\partial x}=0,
\end{equation}%
\begin{equation}\label{eqmotp}
\rho _{0}\frac{\partial u_{x}}{\partial t}=-\frac{\partial }{\partial
x}%
\left[ p_{1}+\frac{B_{0}(x)b_{z}}{4\pi }\right] ,
\end{equation}%
\begin{equation}\label{eqindp}
\frac{\partial b_{z}}{\partial t}+u_{x}\frac{dB_{0}(x)}{dx}%
=-B_{0}(x)\frac{\partial u_{x}}{\partial x},
\end{equation}%
\begin{equation}\label{eqenerp}
\frac{\partial p_{1}}{\partial t}+u_{x}\frac{dp_{0}}{dx}%
=-\gamma p_{0}\frac{\partial u_{x}}{\partial x},
\end{equation}%
where $u_{x}$, $\rho _{1}$, $p_{1}$ and $b_{z}$ denote the
velocity, density, pressure and magnetic field perturbations,
respectively. The solution of the above system is obtained by
using a normal mode decomposition of the perturbation quantities
setting $u_x (x,t)=u(x)e^{i\omega t}$. Then $u(x)$ satisfies (see,
for example, Roberts \cite{rob1981}, \cite{rob3}):
\begin{equation}\label{eqmain}
\frac{d }{d x}\left[ \left( \gamma p_{0}(x) +  \frac{%
B^2_{0}(x)}{4\pi }\right) \frac{%
d u(x)}{d x} \right] + \omega^2 \rho _{0}(x)u(x)=0.
\end{equation}
This equation governs the dynamics of fast magnetosonic waves
propagating across the magnetic field lines in the inhomogeneous
medium (Figure~\ref{figbas}).

The solution of this equation for
different particular equilibrium conditions can be obtained
either analytically or numerically. Equation
(\ref{eqmain}) describes the propagation of a fast wave with speed
$(C_s ^2+V_A ^2)^{1/2}$, where $C_s=(\gamma p_0/\rho _0)^{1/2}$ is
the sound speed and $V_A=(B_0 ^2/4 \pi \rho _0)^{1/2}$ is the
Alfv\'en speed.

We study Eq. (\ref{eqmain}) in accordance with the boundary
conditions:
\begin{equation}\label{bounf}
u(0)=u(L_x)=0,
\end{equation}
corresponding to fast waves bounded by walls located at $x=0$ and
$x=L_x$. These boundary conditions make the spectrum of fast modes
discrete: Eq.~(\ref{eqmain}) has a nontrivial solution only for a
discrete set of frequencies,
\begin{equation}\label{freq}
\omega=\omega _n =\omega _0, \, \omega _1,\ldots.
\end{equation}
In this case the solutions for different physical quantities can
be represented as
\begin{displaymath}
u_{x}=\alpha v(x)\sin (\omega _n t), \hskip 0.2cm \rho _{1}=\alpha
r(x)\cos
(\omega _n t),
\end{displaymath}
\begin{equation}
b_{z}=\alpha h(x)\cos (\omega _n t),
\end{equation}
where the density and magnetic field perturbations are related to
the velocity perturbations through
\begin{equation}
r(x)=\frac{1}{\omega _n}\left( v(x)\frac{d\rho _{0}}{dx}+\rho
_{0}\frac{dv(x)%
}{dx}\right),
\end{equation}
\begin{equation}
h(x)=\frac{1}{\omega _n}\left(
v(x)\frac{dB_{0}}{dx}+B_{0}\frac{dv(x)}{dx} \right).
\end{equation}
Here (and subsequently) we use a subscript $n$ to denote the
frequency of a given standing fast mode.
\section{Swing amplification of Alfv\'en waves}
Now consider Alfv\'en waves that are linearly polarized in the $y$
direction and propagate along the magnetic field (see Figure
\ref{figbas}). In the linear limit these waves are decoupled from
the magnetosonic waves and the equations governing their dynamics
are (see Chen \& Hasegawa \cite{chhas}, Heyvaerts \&
Priest \cite{heypr}):
\begin{equation}
\frac{\partial b_{y}}{\partial t}=B_{0} (x)\frac{\partial
u_{y}}{\partial z},
\end{equation}%
\begin{equation}
\frac{\partial u_{y}}{\partial t}=\frac{%
B_{0}(x)}{4\pi \rho _{0}(x) }\frac{\partial b_{y}}{\partial z}.
\end{equation}
Combining these equations, we obtain the well-known wave equation
governing the propagation of linearly polarized Alfv\'en waves:
\begin{equation}\label{alflin}
\frac{\partial ^2 b_{y}}{\partial t^2}-V_A ^2 (x)\frac{\partial ^2
b_{y}}{\partial z^2}=0,
\end{equation}
where $V_A(x)$ is Alfv\'en speed. It is clear from
Eq.~(\ref{alflin}) that the phase speed of this mode depends on
$x$ parametrically. Therefore, an Alfv\'en wave with a given wave
length propagates with a '{\lq local\rq} characteristic frequency.
Each magnetic flux surface
can evolve independently in this perturbation mode.
\begin{figure*}
\centering
\includegraphics[scale=0.75]{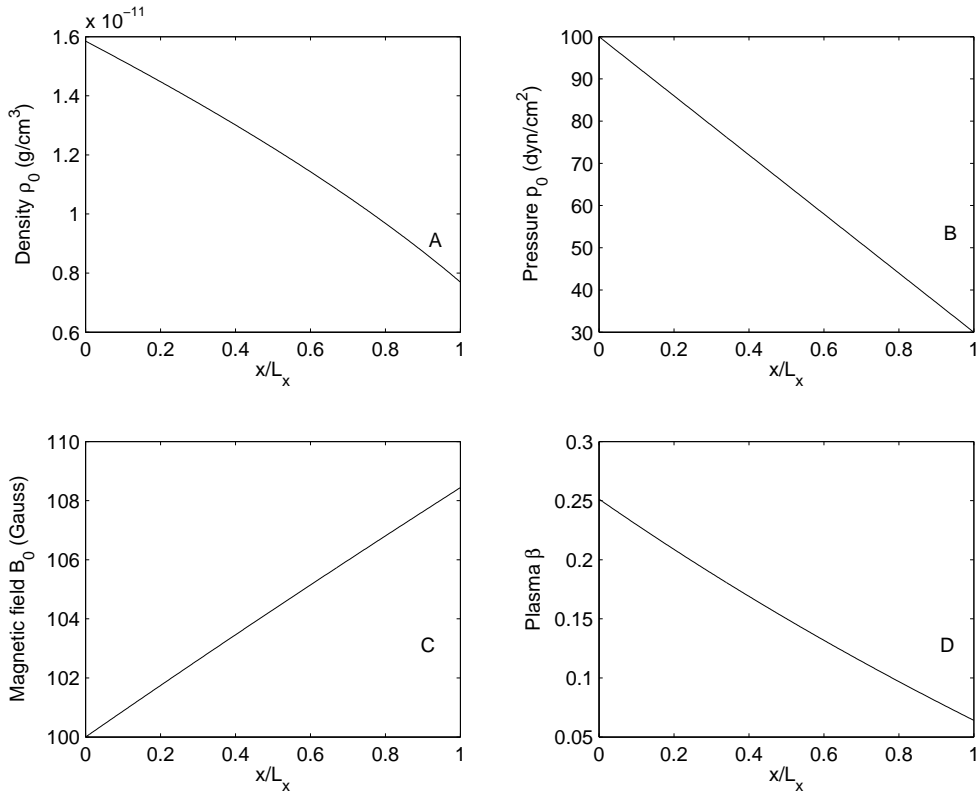}
\caption{The profiles of the equilibrium quantities plotted
against dimensionless $x/L_x$ coordinate:  density (panel A),
pressure (panel B), the magnetic field (panel C) and plasma beta
(panel D) }\label{equil}.
\end{figure*}
\subsection{Propagating Alfv\'en waves}
Let us now address the non-linear action of the fast magnetosonic
waves, considered in the previous section, on Alfv\'en waves.
We study the weakly
non-linear regime. This means that the amplitudes of the fast
magnetosonic waves are considered to be large
enough to produce significant variations of the environment
parameters, which can be felt by propagating Alfv\'en modes, but
too small to affect the Alfv\'en modes themselves.
Hence, the magnetic flux surfaces can still
evolve independently. Therefore, as in paper~I, the non-linear
terms in the equations arising from the advective derivatives
$u_{x}\partial b_{y}/\partial x$ and $\left( \rho _{0}+\rho
_{1}\right) u_{x}\partial u_{y}/\partial x$ are assumed to be
negligible. Under these circumstances the governing set of
equations takes the form (see Paper I):
\begin{equation}\label{alfnon1}
\frac{\partial b_{y}}{\partial t}=\left( B_{0}+b_{z}\right)
\frac{\partial
u_{y}}{\partial z}-\frac{\partial u_{x}}{\partial x}b_{y},
\end{equation}%
\begin{equation}\label{alfnon2}
\left( \rho _{0}+\rho _{1}\right) \frac{\partial u_{y}}{\partial
t}=\frac{%
B_{0}+b_{z}}{4\pi }\frac{\partial b_{y}}{\partial z}.
\end{equation}
These equations describe the parametric influence of fast
magnetosonic waves propagating across the magnetic field on
Alfv\'en waves propagating along the field.
An analytical solution of Eqs. (\ref{alfnon1}) and (\ref{alfnon2})
is possible for a standing pattern of fast magnetosonic
waves, the medium being assumed bounded in the $x$
direction.

Combining Eqs.~(\ref{alfnon1}) and (\ref{alfnon2}) we obtain the
following second order partial differential equation:
\begin{displaymath}
\frac{\partial ^{2}b_{y}}{\partial t^{2}}+\left[ \frac{\partial
u_{x}}{%
\partial x}-\frac{1}{B_{0}+b_{z}}\frac{\partial b_{z}}{\partial
t}\right] \frac{\partial b_{y}}{\partial t} +
\end{displaymath}
\begin{equation}
+\left( \frac{\partial ^{2}u_{x}}{\partial
t\partial x}-\frac{1}{B_{0}+b_{z}}\frac{\partial b_{z}}{\partial
t}\frac{%
\partial u_{x}}{\partial x}\right) b_{y}-\frac{(B_{0}+b_{z})^{2}}{4\pi
\left( \rho _{0}+\rho _{1}\right) }\frac{\partial ^{2}b_{y}}{\partial
z^{2}}%
=0.
\end{equation}
Writing
\begin{equation}
b_{y}=h_{y}(z,t)\exp \left[ -\frac{1}{2}
\int \left( \frac{\partial u_{x}}{\partial
x}-\frac{1}{B_{0}+b_{z}}\frac{\partial b_{z}}{\partial t}\right)
dt\right],
\end{equation}
we obtain
\begin{equation}\label{eqalvin}
\frac{\partial ^{2}h_{y}}{\partial t^{2}}+\frac{1}{2}\left[
S_1 (x,t)-S_2 (x,t)\right ]h_{y}-S_3 (x,t)\frac{\partial
^{2}h_{y}}{\partial z^{2}}=0,
\end{equation}
where,
\begin{equation}\label{s1}
 S_1  = \frac{\partial
^{2}u_{x}}{\partial t\partial
x}+\frac{1}{B_{0}+b_{z}}\frac{\partial ^{2}b_{z}}{\partial t^{2}},
\end{equation}
\begin{equation}\label{s2}
S_2= \frac{1}{\left ( B_0+b_z\right ) ^2}\left (\frac{\partial
b_z}{\partial t}\right ) ^2+\frac{1}{2}\left [ \frac{\partial
u_{x}}{\partial x}+\frac{1}{B_{0}+b_{z}}\frac{\partial
b_{z}}{\partial t}\right ]^2,
\end{equation}
\begin{equation}\label{s3}
 S_3 = \frac{(B_{0}+b_{z})^{2}}{4\pi \left( \rho _{0}+\rho
 _{1}\right)}.
\end{equation}
Finally, applying a Fourier analysis with respect to the $z$
coordinate,
\begin{equation}
h_{y}(z,t)=\int \hat{h}_{y}(k_{z},t)e^{ik_{z}z}dk_{z},
\end{equation}
and neglecting the second and higher order terms in $\alpha$, we obtain the
following Mathieu-type equation:
\begin{equation}\label{alfmain}
\frac{\partial ^{2}\hat{h}_{y}}{\partial
t^{2}}+k_{z}^{2}V_{A}^{2}\left[ 1+\alpha \digamma (x)\cos (\omega
_n t)\right] \hat{h}_{y}=0,
\end{equation}
where
\begin{equation}\label{mathie}
\digamma (x)=2\frac{h(x)}{B_{0}}-\frac{r(x)}{\rho _{0}}-v(x)\frac{%
\omega _n}{2k_{z}^{2}V_{A}^{2}}\frac{1}{B_{0}}\frac{dB_{0}}{dx}.
\end{equation}
It should be noted that the expression (\ref{s2}) for $S_2 (x,t)$
consists only of terms of second order and higher in $\alpha$, and
so can be neglected directly for the case of weakly non-linear
action addressed here.

Equation~(\ref{alfmain}) has a resonant solution when the
frequency of the Alfv\'en mode $\omega _A$ is half of $\omega _n$:
\begin{equation}\label{rescond}
\omega _{A} = k_{z}V_{A}(x)\approx \frac{1}{2}\omega _n.
\end{equation}
This solution can be expressed as
\begin{equation}
{\hat h_y} (k_z,t)={\hat h_y}(k_z,t=0)e^{{{\left |{\delta}\right
|}\over
{2\omega_n}}t}\left
   [{\cos}{{\omega_n}\over 2}t - {\sin}{{\omega_n}\over 2}t \right ],
\end{equation}
where
\begin{equation}\label{delta}
\delta (x)=\alpha k_{z} ^2 V_{A} ^2(x)\digamma (x).
\end{equation}
The solution has a resonant nature within the frequency interval
\begin{equation}\label{resineq}
\left | \omega _A - \frac{\omega _n}{2} \right |<\left |
\frac{\delta}{\omega _n} \right |.
\end{equation}
Similar expressions have been obtained in the Paper~I for a
homogeneous medium. In that case, the Alfv\'en speed is constant
and, therefore, the fast magnetosonic waves amplify the Alfv\'en
waves with the same wavelength everywhere. In the case of an
inhomogeneous Alfv\'en speed, the resonance condition
(\ref{rescond}) implies that the wavelength of the resonant
harmonics of the Alfv\'en waves depends on $x$. This means that
the fast magnetosonic waves now amplify Alfv\'en waves with
different wavelengths (but with the same frequency) in different
magnetic flux surfaces (i.e.,\ different $x$-values).
\subsection{Standing Alfv\'en waves}
When we consider a system that is bounded in the $z$ direction,
the boundary conditions along the $z$ axis introduce an additional
quantization of the wave parameters. In particular, in this case
each spatial harmonic of the Alfv\'en mode can be represented as
\begin{equation}
\hat{h}_{y} ^{m}=\hat{h}_{y}(k_{m},t)\cos
(k_{m}z),
\end{equation}
where $k_{m}=\pi m/L_{z}$ ($m=1,2,\ldots$) and $L_{z}$ is the
characteristic length of the system in the $z$ direction. This
then leads to a further localization of the spatial region where
the swing transfer of wave energy from longitudinal to transversal
oscillations is permitted. The resonant condition (\ref{rescond})
implies that
\begin{equation}\label{rescondm}
k_{m}V_{A}(x_{n,m})\approx \frac{1}{2}\omega _{n}.
\end{equation}
Therefore, the resonant areas are concentrated around the points
$x_{n,m}$ for which condition (\ref{rescondm}) holds. Within these
resonant areas the longitudinal oscillations damp effectively and
their energy is transferred to transversal oscillations with wave
numbers $k_z=k_{m}$ satisfying the resonant conditions. These
resonant areas are localized in space and can be referred to as
regions of {\it swing absorption of the fast magnetosonic
oscillations} of the system. The particular feature of this
process is that the energy transfer of fast magnetosonic waves to
Alfv\'en waves occurs at half the frequency of the fast waves.
\begin{figure*}
\includegraphics[scale=1]{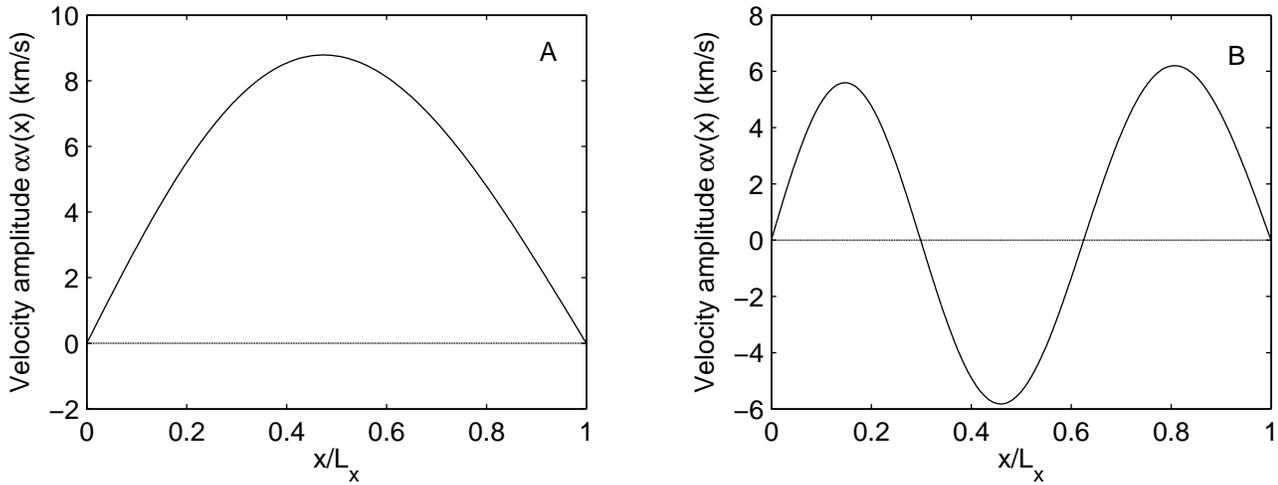}
\caption{Sample solutions of the standing fast magnetosonic modes.
Panel A: zeroth-order harmonic $n=0$, period $T=5.6069$ min.,
$\alpha=0.015$ ; Panel B: second-order harmonic $n=2$, period
$T=1.8514$ min., $\alpha=0.03$}\label{fapa1}.
\end{figure*}

\section{Numerical study of the problem}

In this section we consider in detail the process of swing
absorption of fast waves into Alfv\'en waves. We consider a
numerical study of equation (\ref{eqmain}) subject to the boundary
conditions (\ref{bounf}). We obtained a numerical solutions of
Eqs. (\ref{alfnon1}) and (\ref{alfnon2}) by using the standard
Matlab numerical code for solving sets of ordinary differential
equations. We study, as an example, the case of a polytropic
plasma when both the thermal and magnetic pressures are linear
functions of the $x$ coordinate:
\begin{equation}\label{pprof}
p_{0}=p_{00}+p_{01}\frac{x}{L_x},
\end{equation}
\begin{equation}\label{dprof}
\rho_{0}=C^2\left ( p_{00}+p_{01}\frac{x}{L_x}\right
)^{\frac{1}{\gamma}},
\end{equation}
\begin{equation}\label{bprof}
B_{0}=\sqrt{h_{00}+h_{01}\frac{x}{L_x}},
\end{equation}
where, $p_{00}$, $p_{01}$, $h_{00}$, $h_{01}$ and $C$ are
constants, and $L_{x}$ denotes the length of the system along the
$x$ direction. The pressure balance condition (\ref{balance})
immediately yields
\begin{equation}
p_{01}=-\frac{h_{01}}{8\pi}.
\end{equation}
The solution of the wave equation depends on the values of the
above set of constant parameters. In general, different
equilibrium regimes can be considered including those
corresponding to different limits of the plasma ${\beta}$: $\beta
\ll 1$, $\beta \approx 1$ and $\beta \gg 1$. However, here we
consider the case of the $\beta$ profile shown in Figure
\ref{equil} (panel D).
\begin{table}[h]
\begin{center}
\begin{tabular}{cccc}
\hline \hline $p_{00}$ dyn/cm$^2$&$p_{01}$ dyn/cm$^2$&$h_{00} $
G$^2$&$h_{01}$ G$^2$\\ \hline
100&-70&$10^{4}$&$1.7593 \cdot 10^{3}$\\
\hline
\hline
$L_x$ km & $L_z$ km & $C$ &$\gamma$\\
\hline
15000&$6.5\cdot L_x$&$10^{-6}$&$5/3$\\
\hline \hline
\end{tabular}
\end{center}
\caption{Values of the constant parameters used in the calculation of
our illustrative solutions. The dimension of $C$ is
g$^{1/2}$cm$^{(2-3\gamma)/2\gamma}$/$dyn
^{1/2\gamma}$}\label{tab1}.
\end{table}
The values of all constant parameters are given in
Table~\ref{tab1}. We took arbitrary values of parameters, but they
are somewhat appropriate to the magnetically dominated solar
atmosphere (say the chromospheric network). In Figure \ref{equil}
corresponding equilibrium profiles of the density (panel A),
pressure (panel B), the magnetic field (panel C) and plasma beta
(panel D) are shown. In Figure~\ref{fapa1} we show the profiles of
$\alpha v(x)$ for the standing wave solutions, for two cases with
different modal `wavelength'. Panels A and B, respectively,
correspond to the characteristic frequencies: ~$\omega _0\approx
1.87\cdot 10^{-2}$ s$^{-1}$ (period 5.61 min) and ~$\omega_2
\approx 5.66\cdot 10^{-2}$ s$^{-1}$ (period 1.85 min).

\begin{figure*}
\centering
\includegraphics[scale=0.75]{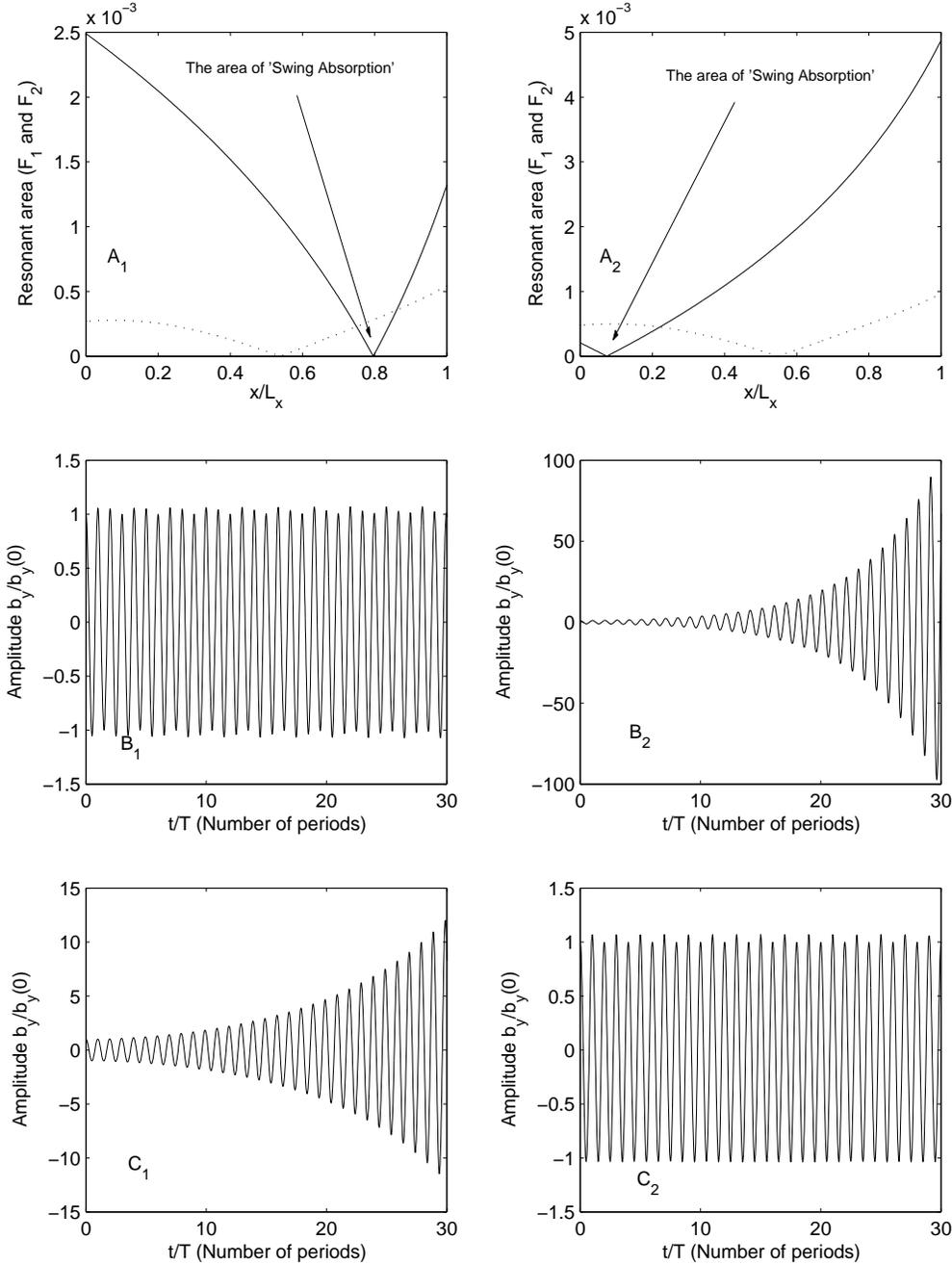}
\caption{ Results of numerical simulations for the standing fast
magnetosonic mode, shown on panel~A in Figure~\ref{fapa1}.
Panel~A$_1$: the curves of $F_1$ (Eq. (\ref{f1})) (solid line) and
$F_2$ (Eq. (\ref{f2})) (dotted line) vs. fractional distance
$x/L_x$ for the zeroth-order standing fast mode $n=0$ and the
standing Alfv\'en mode with $m=3$; Panel B$_1$: The numerical
solution of the Alfv\'en mode $m=3$ amplitude at location
$x/L_x\approx x_{0,4}/L_x=0.0767$; Panel C$_1$: The numerical
solution for mode $m=3$ at $x/L_x\approx x_{0,3}/L_x=0.7967$;
Panel~A$_2$: As in panel~A$_1$ for mode $m=4$; Panel~B$_2$: As in
panel~B$_1$ for mode $m=4$; Panel~C$_2$: As in panel~C$_1$ for
mode $m=4$.}\label{fapb1}
\end{figure*}

\begin{figure*}
\centering
\includegraphics[scale=0.75]{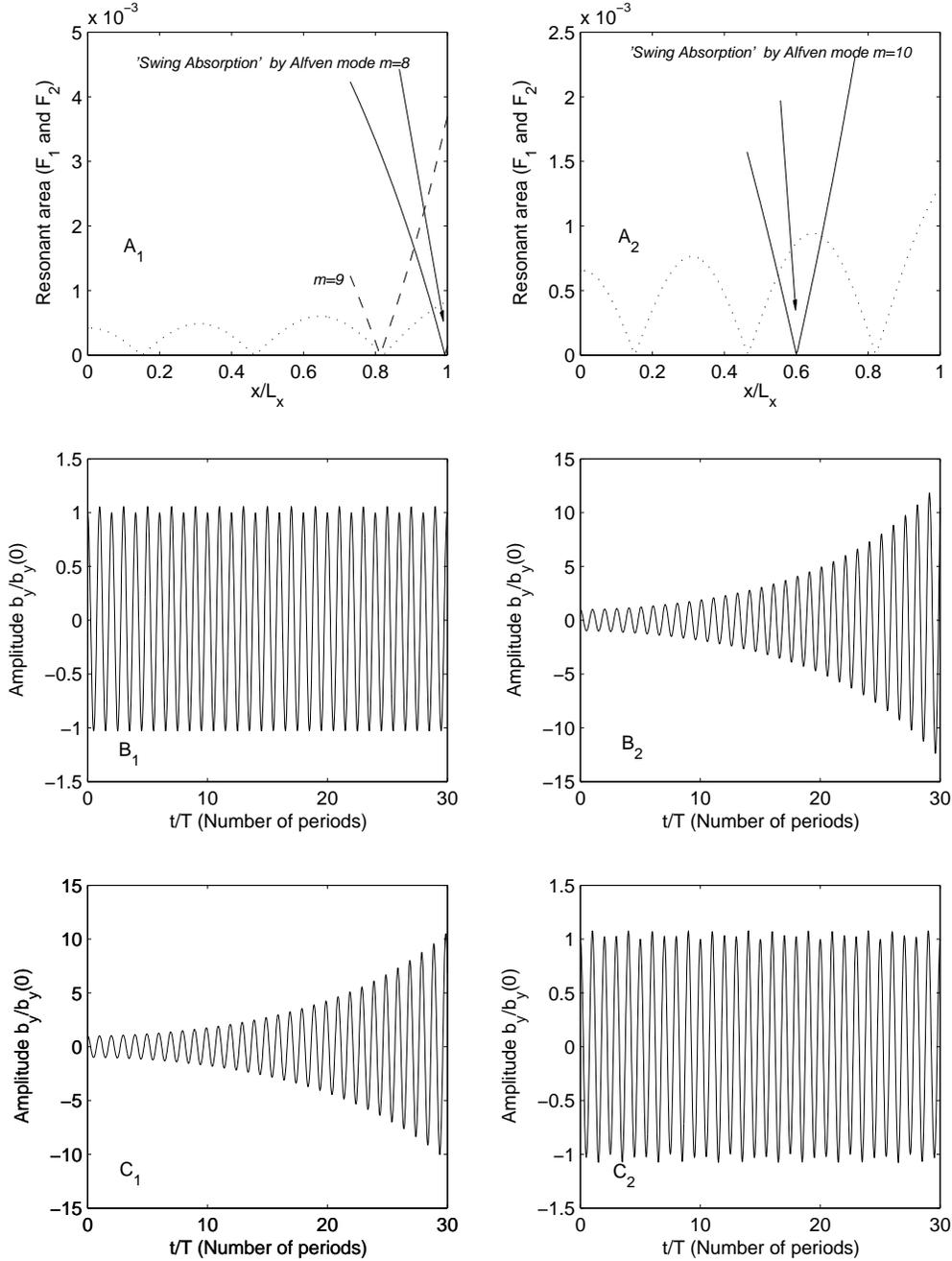}
\caption{ As in Figure \ref{fapb1} for the second-order harmonic
of the fast magnetosonic mode shown on Panel~B in
Figure~\ref{fapa1} ($n=2$) and the standing Alfv\'en with $m=8$
and $m=10$. The numerical solutions respectively are given at
distances $x_1/L_x \approx x_{2,10}/L_x = 0.6000$ (panels B$_1$
and B$_2$), $x_2/L_x \approx x_{2,8}/L_x=0.9933$ (panels C$_1$ and
C$_2$). Dashed line in panel A$_1$ is curve of $F_1$ (\ref{f1})
corresponding to the mode $m=9$.}\label{fapb2}
\end{figure*}

\begin{figure*}
\centering
\includegraphics[scale=0.75]{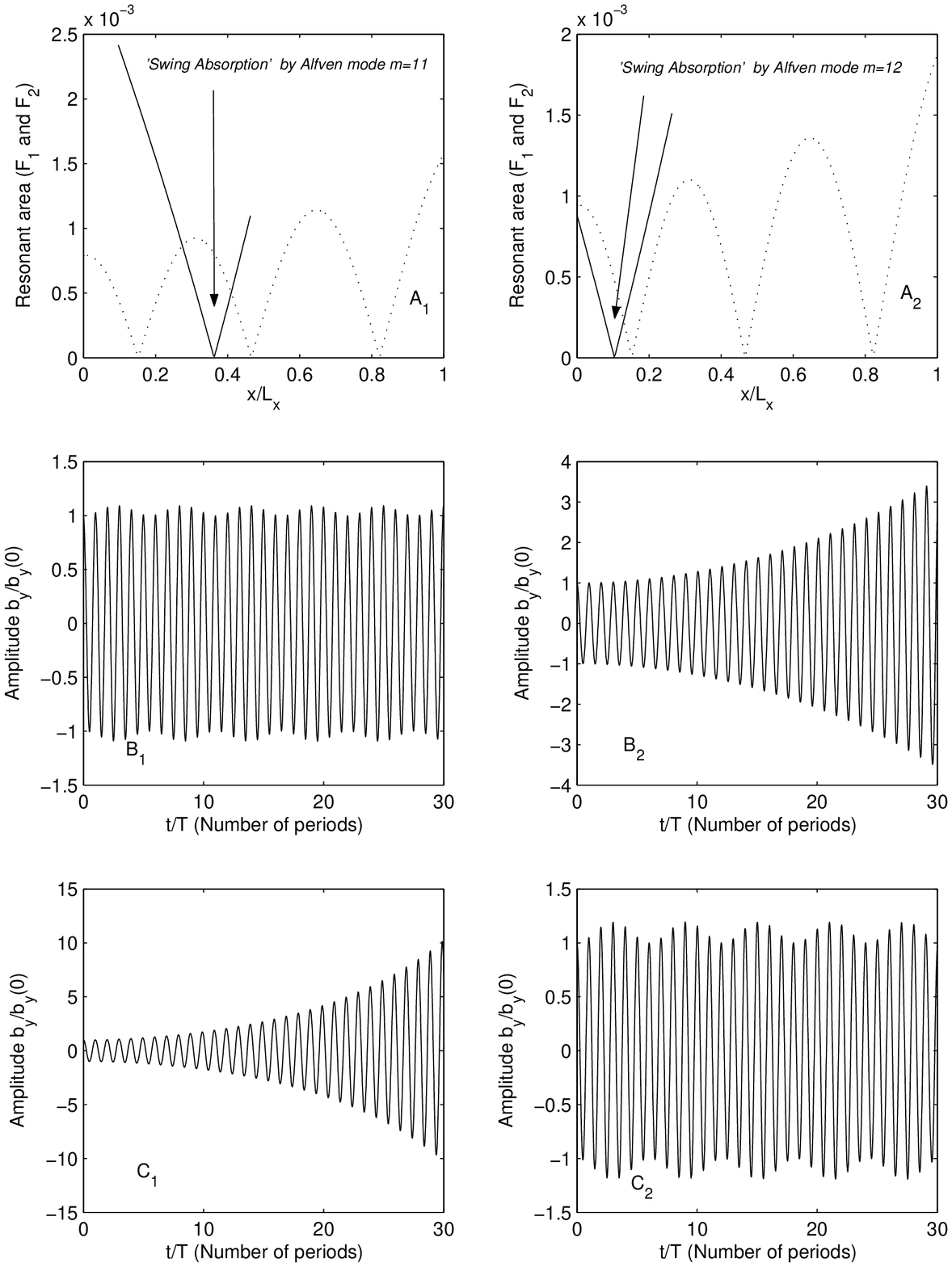}
\caption{ As in Figure~\ref{fapb2} for the standing Alfv\'en modes
with $m=11$ and $m=12$ at respective resonant points $x_1/L_x
\approx x_{2,12}/L_x = 0.1033$ (panels B$_1$ and B$_2$), $x_2/L_x
\approx x_{2,11}/L_x=0.3633$ (panels C$_1$ and
C$_2$).}\label{fapb3}
\end{figure*}

For the configuration described by the equilibrium profiles
(\ref{pprof}) - (\ref{bprof}) the resonant condition
(\ref{rescondm}) yields the areas of swing absorption (located
along the $x$ axis) as the solutions of the following equation:
\begin{equation}
\sqrt{\pi} \omega _{n}C\left(
p_{00}+p_{01}\frac{x_{n,m}}{L_x}\right)
^{\frac{%
1}{2\gamma }}-k_{m}\sqrt{ h_{00}+h_{01}\frac{x_{n,m}}{L_x}} =0.
\end{equation}
In Figure~\ref{fapb1} (panels A$_1$ and A$_2$) we plot
\begin{equation}\label{f1}
F_1=\left |\frac{\omega _n}{2}-k_m V_A \right |
\end{equation}
(solid line) and
\begin{equation}\label{f2}
F_2=\left | \frac{\delta (x)}{\omega _n} \right |
\end{equation}
(dotted line) against the normalized coordinate $x/L_x$. These
curves correspond to the zeroth-order harmonic of the fast
magnetosonic mode shown in Figure~\ref{fapa1} (panel A) and the
standing Alfv\'en mode with wave numbers $m=3$ (panel A$_1$) and
$m=4$ (panel A$_2$).

In order to examine the validity of the approximations we made
during the analysis of the governing equation (\ref{eqalvin}), we
performed a direct numerical solution of the set of equations
(\ref{alfnon1}) - (\ref{alfnon2}) and obtained the following
results. The Alfv\'en mode $m=3$ is amplified effectively close to
the resonant point $x/L_x \approx x_{0,3}/L_x =0.7967$. This is
shown on panel~C$_1$ of Figure~\ref{fapb1}. Far from this resonant
point, the swing interaction is weaker, as at the point
$x/L_x\approx x_{0,4}/L_x=0.0767$ (panel B$_1$). For the Alfv\'en
mode $m=4$ we have the opposite picture: the area of {\lq swing
absorption\rq} is situated around the point $x_{0,4}$ (see panel
B$_2$, Figure~\ref{fapb1}) and the rate of interaction between
modes decreases far from this area, as at point $x_{0,3}$ (panel
C$_2$, Figure~\ref{fapb1}). In these calculations we took $\alpha
= 0.015$.

Similar results are obtained for the fast magnetosonic mode shown
in panel~B of Figure~\ref{fapa1}, corresponding to $\alpha= 0.03$.
In this case, the fast magnetosonic mode effectively amplifies
four different spatial harmonics of Alfv\'en modes, viz.\ $m=8,
10$, $11$ and $12$. There exists one additional resonant point
corresponding to the Alfv\'en mode $m=9$. In panel~A$_1$ of
Figure~\ref{fapb2} we show the curve of $F_1$ for this mode by a
dashed line. But, as seen from this plot, the value of $F_2$, and
consequently the rate of mode amplification $\delta (x)$ in this
resonant point, is very small. Therefore the interaction between
the modes $n=2$ and $m=9$ is negligible for the considered
parameters.

The results of direct numerical calculations for these Alfv\'en
wave harmonics are shown separately in Figures~\ref{fapb2} -
\ref{fapb3}. In Figure~\ref{fapb2} (panel~A$_1$), we show the
curves corresponding to the mode numbers $n=2$, $m=8$. The
calculations were performed at two points $x_1/L_x \approx
x_{2,10}/L_x= 0.60$ (panel B$_1$) and $x_2/L_x\approx
x_{2,8}/L_x=0.9933$ (panel C$_1$). In addition, the results of
calculations for the mode $m=10$, under the same conditions, are
shown in panels~A$_2$, ~B$_2$ and ~C$_2$. It can be seen from
these plots that the $m=8$ Alfv\'en mode is amplified effectively
within the resonant area close to the resonant point $x_{2,8}$
(see panel~C$_1$), which corresponds to the harmonic of the
standing Alfv\'en mode with half the frequency. On the other hand,
in panels B$_1$ we see that the action of the fast mode on the
Alfv\'en mode is almost negligible outside the swing amplification
area. Correspondingly, we get similar results for the Alfv\'en
modes $m=10$ (Figure \ref{fapb2}, panels~A$_2$, B$_2$ and C$_2$),
and for $m=11$ and $m=12$ (Figure \ref{fapb3}) (in the latter case
the calculations were done at points $x_3/L_x\approx
x_{2,12}/L_x=0.1033$ and $x_4/L_x \approx x_{2,11}/L_x = 0.3633$).
Again, all these standing Alfv\'en modes gain their energy from
the fast magnetosonic mode $n=2$ only at the resonant points
$x_{2,10}$ (for $m=10$, panel~B$_2$ in Figure \ref{fapb2}),
$x_{2,11}$ (for $m=11$, panel C$_1$ in Figure \ref{fapb3}) and
$x_{2,12}$ (for $m=12$, panel B$_2$ in Figure \ref{fapb3}),
respectively.

As a conclusion from the above analysis one can claim that the
zeroth-order harmonic of the standing fast magnetosonic mode
propagating across the magnetic field lines, in the system with
characteristic length scales $L_x$ and $L_z=6.5L_z$, can be
effectively absorbed only by the standing Alfv\'en modes with
modal numbers $m=3$ and $m=4$ within the resonant areas,
respectively, around the point of swing absorption $x_{0,3}$ and
$x_{0,4}$. On the other hand, the second-order harmonic of the
standing fast mode $n=2$ is absorbed by four harmonics of the
standing Alfv\'en modes, $m=8, 10, 11,$ and $12$, with respective
locations of the resonant areas of swing absorption at $x_{2,8}$,
$x_{2,10}$, $x_{2,11}$ and $x_{2,12}$. A similar analysis can be
performed for the case of any other equilibrium configuration and
corresponding harmonics of the standing fast magnetosonic modes.

\section{Discussion}

The most important characteristic of swing absorption is that the
velocity polarization of the amplified Alfv\'en wave is strictly
perpendicular to the velocity polarization (and propagation
direction) of fast magnetosonic waves. This is due to the
parametric nature of the interaction. For comparison, the
well-known resonant absorption of a fast magnetosonic wave can
take place only when it does not propagate strictly perpendicular
to the magnetic flux surfaces and the plane of the Alfv\'en wave
polarization. In other words, the energy in fast magnetosonic
waves propagating strictly perpendicular (i.e.\ $k_\bot=0$) to the
magnetic flux surfaces cannot be resonantly {\lq absorbed\rq} by
Alfv\'en waves with the same frequency polarized in the
perpendicular plane. This is because the mechanism of resonant
absorption is analogous to the mechanical pendulum undergoing the
direct action of an external periodic force. This force may
resonantly amplify only those oscillations that at least partly
lie in the plane of force. On the contrary, the external periodic
force acting parametrically on the pendulum length may amplify the
pendulum oscillation in any plane. A similar process occurs when
the fast magnetosonic wave propagates across the unperturbed
magnetic field. It causes a periodical variation of the local
Alfv\'en speed and thus affects the propagation properties of the
Alfv\'en waves. As a result, those particular harmonics of the
Alfv\'en waves that satisfy the resonant conditions grow
exponentially in time. These resonant harmonics are polarized
perpendicular to the fast magnetosonic waves and have half the
frequency of these waves. Hence, for standing fast magnetosonic
waves with frequency $\omega_n$, the resonant Alfv\'en waves have
frequency ${\sim}\omega_n/2$.

In a homogeneous medium all resonant harmonics have the same
wavelengths (see Paper~I). Therefore, once a given harmonic of the
fast and Alfv\'en modes satisfies the appropriate resonant
conditions (Eqs. (23) and (25) in Paper~I), then these conditions
are met within the entire medium. Thus, in a homogeneous medium
the region where fast modes effectively interact with the
corresponding Alfv\'en waves is not localized, but instead covers
the entire system. However, when the equilibrium is inhomogeneous
across the applied magnetic field, the wavelengths of the resonant
harmonics depend on the local Alfv\'en speed. When the medium is
bounded along the unperturbed magnetic field (i.e.\ along the $z$
axis), the resonant harmonics of the standing Alfv\'en waves
(whose wavelengths satisfy condition (\ref{rescondm}) for the
onset of a standing pattern) will have stronger growth rates. This
means that the {\lq absorption\rq} of fast waves will be stronger
at particular locations across the magnetic field. In the previous
section we showed numerical solutions of standing fast
magnetosonic modes for a polytropic equilibrium ($p_0 \sim \rho _0
^\gamma$) in which the thermal pressure and magnetic pressure are
linear functions of $x$. Further, we performed a numerical
simulation of the energy transfer from fast magnetosonic waves
into Alfv\'en waves at the resonant locations, i.e.\ the regions
of {\it swing absorption}.

The mechanism of swing absorption
can be of importance in a veriety of astrophysical situations. Some
possible applications of the mechanism are discussed briefly in the
following subsections.
\subsection{Swing absorption of global fast magnetosonic waves in
the Earth's magnetosphere}
Global standing magnetosonic waves, resulting from the fast
MHD waves being reflected between the bow shock or magnetopause
and a turning point within the planetary magnetosphere, may be
driven by the solar wind (Harrold \& Samson, \cite{har}).
The proposed swing absorption mechanism suggests that these waves may be
{\lq absorbed\rq} by shear Alfv\'en waves with the half frequency,
which may form a standing pattern when they are guided along
magnetic field lines in the magnetosphere-ionosphere system
(reflected by the lower ionosphere at the ends of the magnetic
field lines). They are usually described as field line resonances
and have been observed as regular ultra-low frequency variations
in the magnetic field in the Earth's magnetosphere and F region
flows in the ionosphere (Samson et al. \cite{sam}, Walker et al.
\cite{wal}). This standing Alfv\'en wave pattern may act as an
electromagnetic coupling mechanism between the auroral
acceleration region of the magnetosphere and the ionosphere
(Rankin et al. \cite{ran}). The results obtained here may
be applied to the particular spatial distribution of the
background density and magnetic field
of the Earth's magnetosphere.
\subsection{Absorption of p-modes in sunspots}
Solar $p$-modes interact strongly with sunspots.
Sunspots (and other magnetic field concentrations)
scatter and absorb significantly ingoing acoustic modes (see Braun
et al. \cite{brdulb}, \cite{bretl}; Bogdan et al. \cite{bogdetl};
Zhang \cite{zha}). Understanding the mechanisms leading to the
transformation and damping of waves is thus an important issue in
{\it sunspot seismology} (see Braun \cite{braun}; Cally et al.
\cite{caletl}). The $p$-modes split into fast and slow modes (see, for
example, Spruit \& Bogdan \cite{spru}; Cally \& Bogdan
\cite{calbog}; Bogdan \& Cally \cite{bogdcal}; Cally \cite{call})
and these latter modes have a different nature in a ~high-$\beta$
plasma. In particular, fast modes (or
$\pi$-modes, Cally \& Bogdan \cite{calbog}) behave similarly to
non-magnetized $p$-modes (and they predominantly propagate across
the magnetic field lines), while the nature of slow modes is
closer to that of Alfv\'en modes (they mostly travel along the
field lines). Further, the damping of incident MHD modes,
satisfying local resonant conditions, through the mechanism of
resonant absorption has been intensively studied (see, for example,
Keppens, Bogdan \& Goossens \cite{kepbogdgos}; Stenuit et al.
\cite{sten1}, \cite{sten2}, \cite{sten3}, \cite{sten4}).
The mechanism of {\it swing
absorption} discussed here may also be involved
in the damping of incident
$p$-modes in sunspots, which accordingly satisfy special
resonant conditions (as discussed earlier). Then their
energy can be transferred into Alfv\'en waves
and the amplified Alfv\'en waves may
propagate upwards and carry their energy into the
chromosphere and corona.

\section{Conclusions}

We have shown that swing interaction
(Zaqarashvili \cite{zaq}; Zaqarashvili and Roberts \cite{paper1})
may lead to fast magnetosonic waves, propagating across a
non-uniform equilibrium magnetic field, transforming their
energy into Alfv\'en
waves propagating along the magnetic field.
This process differs from resonant absorption. Firstly, the resonant
Alfv\'en waves have only half the frequency of the incoming fast
magnetosonic waves. Secondly, the velocity of the Alfv\'en waves
is polarized strictly perpendicular to the velocity and
propagation direction of the fast magnetosonic waves. The
mechanism of {\it swing absorption} can be of importance in the
dynamics of the solar atmosphere, the Earth's magnetosphere, and in
astrophysical plasmas generally.

Here we presented the results of a numerical study of
the problem, illustrating the process for two particular harmonics
of the standing fast modes in a specific equilibrium configuration.
Further numerical and
theoretical analysis is needed in order to explore the
proposed mechanism of swing absorption in
the variety of inhomogeneous magnetic structures recorded in observations.
Such tasks are to be the subject of future studies.


\begin{acknowledgements}
This work has been developed in the framework of the pre-doctoral
program of B.M.~Shergelashvili at the Centre for Plasma
Astrophysics, K.U.Leuven (scholarship OE/02/20). The work of T.V
Zaqarashvili was supported by the NATO Reintegration Grant FEL.RIG
980755 and the grant of the Georgian Academy of Sciences. These
results were obtained in the framework of the projects OT/02/57
(K.U.Leuven) and 14815/00/NL/SFe(IC) (ESA Prodex 6). We thank the
referee, Dr R. Oliver, for constructive comments on our paper.
\end{acknowledgements}


\begin{thebibliography}{}
\bibitem[2003]{aschw}
Aschwanden, M. 2003, In `Turbulence, Waves and Instabilities in
the Solar Plasma', Edited by R. Erd\'elyi et al., NATO Science
Series, Kluwer Academic Publishers, p. 215
\bibitem[1997]{bogdcal}
Bogdan, T. J., \& Cally, P. S. 1997, Proc. R. Soc. Lond. A, 453,
943
\bibitem[1993]{bogdetl}
Bogdan, T. J., Brown, T. M., Lites, B.W., \& Thomas, J.H. 1993,
ApJ, 406, 723
\bibitem[1995]{braun}
Braun, D. C. 1995, In Proc. GONG '94 `Helio- and Astero-Seismology
from the Earth and Space', Edited by R. K. Ulrich, E. J. Rhodes
Jr. and W. Dappen, Astronomical Society of the Pacific, San
Francisco, California, Vol. 76, p. 250
\bibitem[1987]{brdulb}
Braun, D. C., Duvall, T. L., \& LaBonte, B. J. 1987, ApJ, 319, L27
\bibitem[1992]{bretl}
Braun, D. C., Duvall, T. L., LaBonte, B. J., Jefferies, S.M.,
Harvey, J. W., \& Pomerantz, M. A. 1992, ApJ, 391, L113
\bibitem[2000]{call}
Cally, P. S. 2000, Solar Phys., 192, 395
\bibitem[1993]{calbog}
Cally, P. S., \& Bogdan, T. J. 1993, ApJ, 402, 732
\bibitem[2003]{caletl}
Cally, P. S., Crouch, A. D., \& Braun D.C. 2003, MNRAS, 346, 381
\bibitem[1996]{charots}
Chagelishvili, G. D., Rogava, A. D., \& Tsiklauri, D. G. 1996,
Phys. Rev E, 53, 6028
\bibitem[1974]{chhas}
Chen, L., \& Hasegawa, A. 1974, Phys. Fluids, 17, 1399
\bibitem[1985]{craik}
Craik, A. D. D. 1985, {\it Wave Interactions and Fluid Flows},
Cambridge University Press
\bibitem[1962]{galor}
Galeev, A. A., \& Oraevski, V. N. 1962, Sov. Phys. Dokl., 7, 998
\bibitem[1965]{gollin}
Goldreich, P., \& Linden-Bell, D. 1965, MNRAS, 130, 125
\bibitem[1991]{goos}
Goossens, M. 1991, In `Advances in Solar System -
Magnetohydrodynamics', Edited by E.R. Priest and A.W. Hood,
Cambridge University Press, p. 137
\bibitem[1992]{har}
Harrold, B. G., \& Samson, J. C. 1992, Geophys. R. L., 19, 1811
\bibitem[1983]{heypr}
Heyvaerts, J., \& Priest, E. R. 1983, A\&A, 117, 220
\bibitem[1987]{holl}
Hollweg, J. V. 1987, ApJ, 317, 514
\bibitem[1978]{ion}
Ionson, J. A. 1978, ApJ, 226, 650
\bibitem[1994]{kepbogdgos}
Keppens, R., Bogdan, T. J., \& Goossens, M. 1994, ApJ, 436, 372
\bibitem[1887]{kel}
Lord Kelvin (W. Thomson) 1887, Phil. Mag., 24, Ser. 5, 188
\bibitem[1997]{nakrobmur}
Nakariakov, V. M., Roberts, B., \& Murawski, K. 1997, Sol. Phys.,
175, 93
\bibitem[1995]{ofm}
Ofman, L., \& Davila, J. M. 1995, J. Geophys.Res., 100, 23427
\bibitem[1983]{ora}
Oraevski, V. N. 1983, in {\it Foundations of Plasma Physics},
edited by A. A. Galeev and R. Sudan, Nauka, Moscow
\bibitem[2002]{poed}
Poedts, S. 2002, In Proc. Euroconference and IAU Colloquium 188
`Magnetic Coupling of the Solar Atmosphere', Santorini, Greece,
ESA SP-505, p. 273
\bibitem[1989]{poegok}
Poedts, S., Goossens, M., \& Kerner, W. 1989, Sol. Phys., 123, 83
\bibitem[1982]{raerob}
Rae, I. C., \& Roberts, B. 1982, MNRAS, 201, 1171
\bibitem[1993]{ran}
Rankin, R., Harrold, B.G., Samson, J.C., \& Frycz, P. 1993, J.
Geophys. R., 98, 5839
\bibitem[1981]{rob1981}
Roberts, B. 1981, Solar Phys., 69, 27
\bibitem[1991]{rob3}
Roberts, B. 1991, In `Advances in Solar System
Magnetohydrodynamics', Edited by E.R. Priest and A.W. Hood,
Cambridge University Press, p. 105
\bibitem[2002]{rob2}
Roberts, B. 2002, In `Solar Variability: From Core to Outer
Frontiers', Prague, Czech Republic, 9-14 September 2002 (ESA
SP-506), p. 481
\bibitem[2004]{rob1}
Roberts, B. 2004, In Proc. SOHO 13 `Waves, Oscillations and
Small-Scale Transient Events in the Solar Atmosphere: A Joint View
from SOHO and TRACE', Palma de Mallorca, Spain, (ESA SP-547), p. 1
\bibitem[2003]{robnak}
Roberts, B., \& Nakariakov, V. M. 2003, In `Turbulence, Waves and
Instabilities in the Solar Plasma', Edited by R. Erd\'elyi et al.,
NATO Science Series, Kluwer Academic Publishers, p. 167
\bibitem[2000]{ropom}
Rogava, A. D., Poedts, S., \& Mahajan S. M. 2000, A\&A, 354, 749
\bibitem[1977]{ryu}
Ryutova, M. P. 1977, In Proc. of XIII Int. Conf. on Ionized Gases,
Berlin, ESA, p. 859
\bibitem[1969]{saggal}
Sagdeev, R. Z., \& Galeev A. A. 1969, {\it Nonlinear Plasma
Theory}, Benjamin, New York
\bibitem[1991]{sam}
Samson, J. C., Greenwald, R. A., Ruohoniemi, J. M., Hughes, T. J.,
\& Wallis, D. D. 1991, Can. J. Phys., 69, 929
\bibitem[1992]{spru}
Spruit, H. C., \& Bogdan, T. J. 1992, ApJ, 391, L109
\bibitem[1993]{sten1}
Stenuit, H., Poedts, S., \& Goossens, M. 1993, Solar Phys., 147,
13
\bibitem[1995]{sten2}
Stenuit, H., Erd\'elyi, R., \& Goossens, M. 1995, Solar Phys.,
161, 139
\bibitem[1998a]{sten3}
Stenuit, H., Keppens, R., \& Goossens, M. 1998, A\&A, 331, 392
\bibitem[1998b]{sten4}
Stenuit, H., Tirry, W. J., Keppens, R., \& Goossens, M. 1998,
A\&A, 342, 863
\bibitem[1992]{wal}
Walker, A. D. M., Ruohoniemi, J. M., Baker, K. B., Greenwald, R.
A., \& Samson, J. C. 1992, J. Geophys. Res., 97, 12187
\bibitem[2001]{zaq}
Zaqarashvili, T. V. 2001, ApJ Letters, 552, L81
\bibitem[2002a]{paper1}
Zaqarashvili, T. V., \& Roberts, B. 2002a, Phys. Rev. E, 66,
026401 (Paper I)
\bibitem[2002b]{zaq0}
Zaqarashvili, T. V., \& Roberts, B. 2002b, In `Solar Variability:
From Core to Outer Frontiers', Prague, Czech Republic, 9-14
September 2002 (ESA SP-506), p. 79
\bibitem[2002]{zaq2}
Zaqarashvili, T. V., Oliver, R., \& Ballester, J. L. 2002, ApJ,
569, 519
\bibitem[2004]{zaq3}
Zaqarashvili, T. V., Oliver, R., \& Ballester, J. L. 2004, In
Proc. SOHO 13 `Waves, Oscillations and Small-Scale Transient
Events in the Solar Atmosphere: A Joint View from SOHO and TRACE',
Palma de Mallorca, Spain, (ESA SP-547), p. 193
\bibitem[1997]{zha}
Zhang, H. 1997, ApJ, 479, 1012
\end{thebibliography}
\end{document}